\documentclass[prd,onecolumn,showpacs,floatfix,superscriptaddress,nofootinbib]{revtex4}
\usepackage{graphicx}
\usepackage{epsfig}
\usepackage{bm}
\usepackage{amssymb}
\usepackage{float}
\usepackage{amsmath}
\usepackage{dcolumn}
\usepackage{cancel}
\usepackage[colorlinks]{hyperref}
\usepackage[usenames,dvipsnames]{color}
\hypersetup{
     breaklinks=true,
    pdfstartview={FitH},    
    colorlinks=true,       
    linkcolor=blue,          
    citecolor=red,        
    filecolor=magenta,      
    urlcolor=blue,           
    anchorcolor=green,      
    linktocpage=true
}

\def\doi{http://doi.org}

\def\be{\begin{equation*}}
\def\ee{\end{equation*}}

\begin{document}

\title{Perturbative and nonperturbative quasinormal modes of 4D Einstein-Gauss-Bonnet black holes}


\author{Almendra Arag\'{o}n}
\email{almendra.aragon@mail.udp.cl} \affiliation{Facultad de
Ingenier\'{i}a y Ciencias, Universidad Diego Portales, Avenida Ej\'{e}rcito
Libertador 441, Casilla 298-V, Santiago, Chile.} 

\author{Ram\'{o}n B\'{e}car}
\email{rbecar@uct.cl}
\affiliation{Departamento de Ciencias Matem\'{a}ticas y F\'{\i}sicas, Universidad Catolica de Temuco}

\author{P. A. Gonz\'{a}lez}
\email{pablo.gonzalez@udp.cl} \affiliation{Facultad de
Ingenier\'{i}a y Ciencias, Universidad Diego Portales, Avenida Ej\'{e}rcito
Libertador 441, Casilla 298-V, Santiago, Chile.}

\author{Yerko V\'asquez}
\email{yvasquez@userena.cl}
\affiliation{Departamento de F\'isica y Astronom\'ia, Facultad de Ciencias, Universidad de La Serena,\\
Avenida Cisternas 1200, La Serena, Chile.}

\date{\today }

\begin{abstract}
We study the propagation of probe scalar fields in the background of 4D Einstein-Gauss-Bonnet black holes with anti-de Sitter (AdS) asymptotics and calculate the quasinormal modes. Mainly, we show that the quasinormal spectrum consists of two different branches, a branch perturbative in the Gauss–Bonnet coupling constant $\alpha$ and another branch nonperturbative in $\alpha$.
The perturbative branch consists of complex quasinormal frequencies that approximate to the quasinormal frequencies of the Schwarzschild AdS black hole in the limit of a null coupling constant. On the other hand, the nonperturbative branch consists of purely imaginary frequencies and is characterized by the growth of the imaginary part when $\alpha$ decreases, diverging in the limit of null coupling constant, therefore they do not exist for the Schwarzschild AdS black hole. Also, we find that the imaginary part of the quasinormal frequencies is always negative for both branches; therefore, the propagation of scalar fields is stable in this background.

\end{abstract}


\maketitle
\flushbottom

\tableofcontents

\newpage

\section{Introduction}

The $4D$ Einstein-Gauss-Bonnet (EGB) gravity has been recently reformulated as the limit $D \rightarrow 4$  of their higher dimensional version when the coupling constant is rescaling as $\alpha \rightarrow \frac{\alpha}{D-4}$ \cite{Glavan:2019inb}. Thus,
the Gauss-Bonnet term shows a nontrivial contribution to the gravitational dynamics. The theory preserves the number of degrees of freedom and remains free from the Ostrogradsky instability. Also, the new theory
has stimulated a series of recent research works concerning to
black holes solutions and the properties of the novel 4D EGB theory, for instance, spherically symmetric black hole solutions were discovered \cite{Glavan:2019inb}, generalizing the Schwarzschild black holes and are also free from singularity. Additionally, charged black holes in AdS spacetime \cite{Fernandes:2020rpa}, radiating black holes solutions \cite{Ghosh:2020syx} and an exact charged black hole surrounded by clouds of string was investigated \cite{Singh:2020nwo}. The generalization of these static black holes to the rotating case was also addressed \cite{Wei:2020ght}. On the other hand, regular black holes and  the generalization of the BTZ solution in the presence of higher curvature (Gauss-Bonnet and Lovelock) corrections of any order was found in Refs. \cite{Kumar:2020uyz} and \cite{Konoplya:2020ibi}, respectively.  Also, a $4D$ Einstein-Lovelock theory was formulated and black hole solutions were studied in \cite{Konoplya:2020qqh, Konoplya:2020der}. The interesting physical properties of the black holes in this novel 4D Einstein-Gauss-Bonnet gravity has been investigated such as their thermodynamics \cite{ Hegde:2020xlv,HosseiniMansoori:2020yfj,Wei:2020poh,Singh:2020xju,EslamPanah:2020hoj}, Hawking radiation and greybody factors  \cite{Zhang:2020qam, Konoplya:2020cbv}, quasinormal modes and stability \cite{Konoplya:2020bxa,Konoplya:2020juj, Mishra:2020gce,Churilova:2020aca,Zhang:2020sjh}, geodesics motion and shadow \cite{Guo:2020zmf,Heydari-Fard:2020sib,Roy:2020dyy}, electromagnetic radiation from a thin accretion disk from spherically symmetric black holes \cite{Liu:2020vkh}, among others. However, recent works has raised criticisms about the approach applied in Ref. \cite{Glavan:2019inb}, that arises from the idea of defining a theory from a set of solutions that are obtained by the limit $D \rightarrow 4$ of the $D$-dimensional EGB theory, and there is an active debate on its validity, see for instance \cite{Gurses:2020ofy, Mahapatra:2020rds, Shu:2020cjw, Tian:2020nzb}. However, in Refs. \cite{Lu:2020iav, Fernandes:2020nbq, Hennigar:2020lsl} have been proposed other approaches to obtain a well defined $D \rightarrow 4$ limit of EGB theory, and an action with a set of field equations were found, by using dimensional reduction methods  \cite{Mann:1992ar, Kobayashi:2020wqy}.
The resulting theory corresponds to a scalar-tensor theory of the Horndeski type. It was shown that all the solutions found in the original paper on $4D$ EGB theory \cite{Glavan:2019inb} are also solutions of the new formulation of the theory. In particular, the spherically symmetric Schwarzschild-like solution generated by this theory coincides with the metric of the $D \rightarrow 4$ limit of the $D$-dimensional EGB theory.
\\

In the context of the detection of gravitational waves \cite{Abbott:2016blz}, the quasinormal modes (QNMs) and quasinormal frequencies (QNFs) are important \cite{Regge:1957td,Zerilli:1971wd, Kokkotas:1999bd, Nollert:1999ji, Konoplya:2011qq}. Despite the detected signal is consistent with the Einstein gravity \cite{TheLIGOScientific:2016src}, there are possibilities for alternative theories of gravity due to the large uncertainties in mass and angular momenta of the ringing black hole \cite{Konoplya:2016pmh}. It have been shown that the spectrum of QNMs of theories with higher curvature corrections, such as the Einstein-Gauss Bonnet  gravity consists of two different branches \cite{Konoplya:2017ymp, Gonzalez:2017gwa, Grozdanov:2016fkt, Grozdanov:2016vgg, Gonzalez:2018xrq}. One of them has an Einsteinian limit when the Gauss-Bonnet coupling constant $\alpha$ tends to zero, while the other consists from purely imaginary modes of which the damping rate is increasing when $\alpha$ decreases, these modes are qualitatively different from their Einsteinian analogues and they do not exists in the limit $\alpha=0$. This branch is, thereby, nonperturbative in $\alpha$ \cite{Gonzalez:2017gwa} {\footnote{Calling nonperturbative to this branch could sound inappropriate 
because it is derived by solving the linearized (perturbative) scalar
equation.}}.

The phenomena of nonperturbative modes seems to be general and independent on the asymptotic behavior of a black hole, topology of the event horizon, spin of the fields under consideration, and, possibly, even of the particular form of the higher curvature corrections to the General Relativity (GR). Thus, nowadays the study of nonperturbative modes has been a subject of interest, due to they may lead to a profile qualitatively different to the gravitational ringdown. On the other hand, from the gauge/gravity duality point of view, these modes lead to the eikonal instability of Gauss-Bonnet black holes at some critical values of coupling constant,  
so that they determine possible constrains on holographic applicability of the black holes backgrounds. Moreover, it is worth to mentioning that the new nonperturbative modes were found for several quite different situation such  as 
the fourth order in curvature theory \cite{Grozdanov:2016fkt}, asymptotically flat  black holes \cite{Gleiser:2005ra, Dotti:2005sq, Takahashi:2010ye} 
and black branes.\\

In this work we consider 4D Einstein-Gauss-Bonnet black holes with anti-de Sitter (AdS) asymptotics and we study the propagation of scalar fields in such backgrounds, in order to show the existence of nonperturbative QNMs for this kind of theories. We obtain the QNFs numerically by using the  pseudospectral Chebyshev method \cite{Boyd} 
which is an effective method to find high overtone modes, and that has been applied for instance in Refs. \cite{Finazzo:2016psx, Gonzalez:2017shu}. In spite of the criticisms on the original $4D$ EGB, it is important to emphasize that the spherically symmetric Schwarzschild-like solution obtained in the $D \rightarrow 4$ limit of the $D$-dimensional EGB theory, is also a solution of theories formulated with a well defined limit $D \rightarrow 4$ of EGB theory. Furthermore, it is worth to noting that these black holes are also solutions of the semi classical Einstein equation with Weyl anomaly \cite{Cai:2009ua} and for a toy model of Einstein gravity with a Gauss-Bonnet classically ``entropic'' term mimicking a quantum correction \cite{Cognola:2013fva}. Therefore, it is worthwhile to perform a study of the physical properties of these black holes, such as the propagation of matter field outside the event horizon.
The QNFs  of scalar, electromagnetic and gravitational perturbations for this background in asymptotically flat spacetime were obtained recently in Ref. \cite{Konoplya:2020bxa}, and it was shown that when the coupling constant is positive, the black hole is gravitationally unstable unless the coupling constant is small enough ($0 < \alpha \lesssim 0.15$). The instability develops at high multipole number $\ell$, and therefore is known as {\it eikonal} instability. Also, the negative coupling constant allows for a stable black-hole solution up to relatively large absolute values of $\alpha$ ($0 > \alpha \gtrsim -2.0$).  The QNFs of Dirac's field, was studied in Ref. \cite{Churilova:2020aca}, and it was shown that the real part of the QNFs is considerably increased, while the damping rate is usually decreasing when the coupling constant increased. Here, besides the perturbative modes, we will find nonperturbative modes in $\alpha$. When $\alpha = 0$ the metric corresponds to the  Schwarzschild AdS  black hole and the QNMs for this geometry were calculated in Ref. \cite{Horowitz:1999jd}, where the approach to thermal equilibrium was established, and previously in Ref. \cite{Chan:1996yk}.\\

The manuscript is organized as follows: In Sec. \ref{background} we give a brief review of the 4D Einstein-Gauss-Bonnet gravity. In Sec. \ref{QNM}, we study the scalar field stability and calculate numerically the QNFs of scalar field perturbations by using the spectral method. Finally, our conclusions are in Sec. \ref{conclusion}.

\section{ Einstein Gauss-Bonnet Black hole in four dimensional AdS spacetime}
\label{background}
The Lagrangian of the $D$-dimensional Einstein-Maxwell-Gauss-Bonnet theory with the coupling constant re-scaled by $\alpha \rightarrow\frac{\alpha}{D-4}$, is given by the relation \cite{Fernandes:2020rpa}
\begin{equation}
\mathcal{L}=R-2 \Lambda+\frac{\alpha}{D-4}\mathcal{G}-F^{\mu\nu}F_{\mu\nu} \, ,
\end{equation}
where $\Lambda=-\frac{(D-1)(D-2)}{2 l^2} $ is the cosmological constant , $\mathcal{G}= R^2- 4R_{\mu \nu} R^{\mu \nu} +R_{\mu \nu \rho \sigma} R^{\mu \nu \rho \sigma}$ is the Gauss-Bonnet term and $F_{\mu \nu}$ is the electromagnetic field tensor.
The solutions for a static and spherically symmetric ansatz in an arbitrary number of dimensions $D\geq 5$, has the form
\begin{equation}
ds^2=-f(r) dt^2+f(r)dr^2+ r^2 d\Omega^2_{D-2} \,,
\end{equation}
where $d\Omega^2_{D-2}$ corresponds to $(D-2)$-dimensional hypersurface. Then, following the prescription given in \cite{Glavan:2019inb} and taking the limit $D\rightarrow4$ it is possible to obtain the exact solution representing the 4D Einstein-Maxwell Gauss-Bonnet black hole \cite{Fernandes:2020rpa}:
\begin{equation}
f(r)=1+\frac{r^2}{2\alpha}\left(1\pm\sqrt{1+4\alpha\left(\frac{2M}{r^3}-\frac{Q^2}{r^4}-\frac{1}{l^2}\right)}\right) \,,
\end{equation}
where $M$ is the mass of black hole and $Q$ is its electric charge. From now on we will consider the uncharged version $Q=0$ of the black hole metric:
\begin{equation}  \label{metrica}
f(r)=1+\frac{r^2}{2\alpha}\left(1\pm\sqrt{1+4\alpha\left(\frac{2M}{r^3}-\frac{1}{l^2}\right)}\right) \,.
\end{equation}
Of the two branches of solution we are interested in the negative branch because it is the most physically interesting one; by taking appropriate limits it is possible to recover some special cases, for instance, when $\alpha \rightarrow 0$, 
 Schwarzschild AdS (SAdS) black hole, and AdS spacetime ($M=0$) 
when $0\leq\alpha\leq\frac{l^2}{4}$ and for null cosmological constant the seminal result found in \cite{Glavan:2019inb} with the coupling parameter $\alpha>0$. It is worth to mention that similar metrics were found previously in the context of quantum corrections to gravity \cite{Tomozawa:2011gp, Cai:2009ua, Cognola:2013fva}.

\begin{figure}[h!] 
\begin{center}
\includegraphics[width=0.42\textwidth]{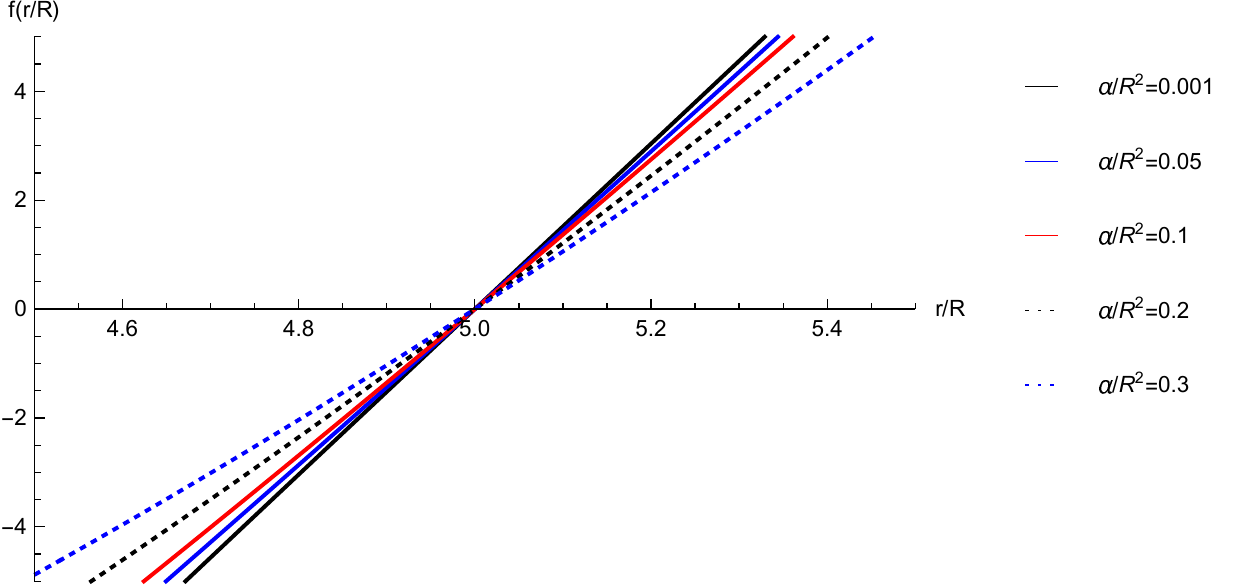}
\end{center}
\caption{The behavior of the metric function $f(r/R)$ as a function of $r/R$ for different values of the parameter $\alpha/R^2$ with $r_H/R=5$,
for the region near the event horizon $r_H/R$.} 
\label{fr}
\end{figure}

The black hole horizon $r_H$ corresponds to the largest root of $f(r) = 0$. In Fig.\ref{fr} we show the behavior of $f(r)$ of different values of $\alpha/R^2$. 
It is convenient to measure all the quantities in the units of the same dimension, so we express $M$ as a function of the event horizon $r_H$:
\begin{equation}
M=-\frac{\Lambda r_H^3}{6}+ \frac{r_H}{2}+\frac{\alpha}{2 r_H} \,,
\end{equation}
where the cosmological constant $\Lambda=-\frac{3}{l^2}$ can be expressed in terms of the AdS radius $R$, which is defined by $f(r \rightarrow \infty)=r^2/R^2$, as
\begin{equation}
\Lambda= -\frac{3(R^2- \alpha)}{R^4} \,.
\end{equation}

\section{Scalar field perturbations}
\label{QNM}

\begin{figure}[h!]
\begin{center}
\includegraphics[width=0.42\textwidth]{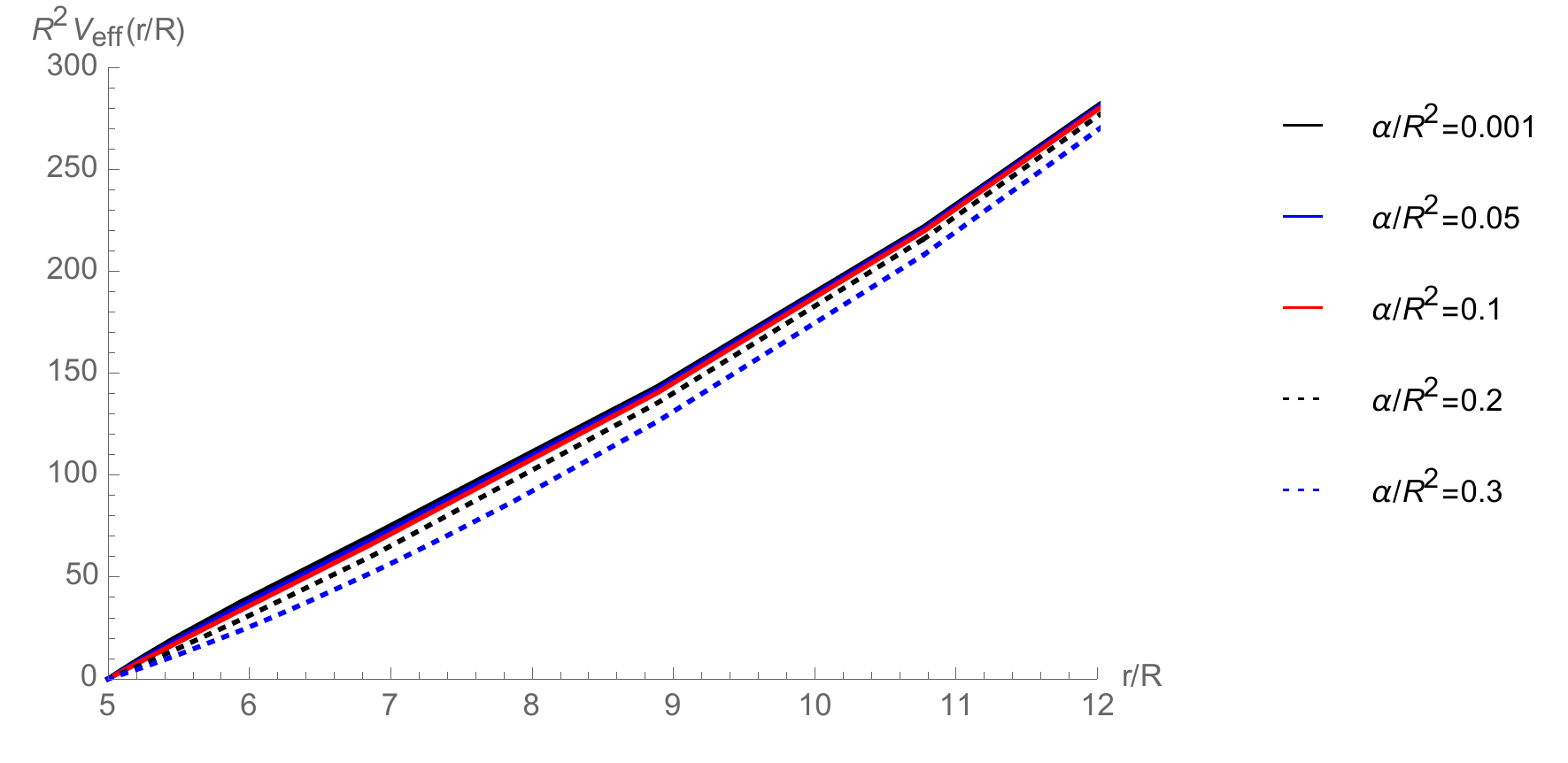}
\end{center}
\caption{The behavior of the effective potential  $R^2 V_{eff}$ as a function of $r/R$ for different values of the parameter $\alpha/R^2$ with $r_H/R=5$, $\ell=0$, and $m R=0.1$.} 
\label{Potential}
\end{figure}

The QNMs of scalar perturbations in the background of  the metric (\ref{metrica}) are given by the scalar field solution of the Klein-Gordon equation
\begin{equation}
\frac{1}{\sqrt{-g}}\partial _{\mu }\left( \sqrt{-g}g^{\mu \nu }\partial_{\nu } \varphi \right) =m^{2}\varphi \,,  \label{KGNM}
\end{equation}%
with suitable boundary conditions for a black hole geometry. In the above expression $m$ is the mass 
of the scalar field $\varphi $. Now, by means of the following ansatz
\begin{equation}
\varphi =e^{-i\omega t} R(r) Y(\Omega) \,,\label{wave}
\end{equation}%
the Klein-Gordon equation reduces to
\begin{equation}
f(r)R''(r)+\left(  f'(r)+2\frac{f(r)}{r} \right)R'(r)+\left(\frac{\omega^2}{f(r)}-\frac{\ell (\ell+1) }{r^2}-m^{2}\right) R(r)=0\,, \label{radial}
\end{equation}%
where $\ell=0,1,2,...$ represents the azimuthal quantum number and the prime denotes the derivative with respect to $r$.
Now, defining $R(r)=\frac{F(r)}{r}$
and by using the tortoise coordinate $r^*$ defined by
$dr^*=\frac{dr}{f(r)}$,
 the Klein-Gordon equation can be written as a one-dimensional Schr\"{o}dinger equation
 \begin{equation}\label{ggg}
 \frac{d^{2}F(r^*)}{dr^{*2}}-V_{eff}(r)F(r^*)=-\omega^{2}F(r^*)\,,
 \end{equation}
 with an effective potential $V_{eff}(r)$, which is parametrically thought as $V_{eff}(r^*)$, given by
  \begin{equation}\label{pot}
 V_{eff}(r)=f(r) \left(\frac{f'(r)}{r}+\frac{\ell (\ell+1)}{r^2} + m^2 \right)~.
 \end{equation}
The effective potential diverges at spatial infinity and it is positive definite everywhere outside the event horizon, see Fig. \ref{Potential}. Therefore, we will consider as a boundary condition that the scalar field vanishes at the asymptotic region ({\it Dirichlet boundary condition}).

\newpage

\subsection{Scalar field stability with Dirichlet boundary condition}

In order to know about the stability of the propagation of scalar fields, we follow a general argument given in Ref. \cite{Horowitz:1999jd}.
So, by defining
\begin{equation}
\psi(r)=e^{i\omega r^{\ast}} F(r)\,,
\end{equation}
and inserting this expression in the Schr\"odinger like equation \eqref{ggg} yields
\begin{equation}\label{KleinFink}
\frac{d}{dr}(f(r)\frac{d\psi(r)}{dr})-2i\omega \frac{d\psi(r)}{dr}-\frac{V_{eff}(r)}{f(r)}\psi(r)=0\,.
\end{equation}
Then, multiplying Eq. (\ref{KleinFink}) by $\psi^{\ast}$ and performing integrations by parts, and using the Dirichlet boundary condition for the scalar field at spatial infinity, one can obtain the following expression
\begin{equation}\label{relacion}
\int _{r_{+}}^{\infty}dr \left( f(r) \left|  \frac{d\psi}{dr}\right|^2+\frac{V_{eff}(r)}{f(r)} \left| \psi \right| ^2 \right)=-\frac{\left|\omega \right|^2 \left| \psi (r=r_H)\right| ^2}{Im(\omega)}\,.
\end{equation}
In general, the QNFs are complex, where the real
part represents the frequency of the oscillation and the imaginary part describes the rate at which this oscillation is damped, with the stability of the scalar field being guaranteed if the imaginary part is negative. The potential (\ref{pot})
 is positive outside the horizon and
then the left hand side of \eqref{relacion} is strictly positive, which demand that $Im (\omega)<0$, and then we conclude that the stability of the propagation of a scalar field respecting Dirichlet boundary conditions is stable.

\subsection{Numerical analysis}

In this section we will solve numerically the differential equation (\ref{radial}) in order to compute the QNFs for the black hole described by the metric by using the pseudospectral Chebyshev method, see for instance \cite{Boyd}. 
First, under the change of variable $y=1-r_H/r$ the radial equation (\ref{radial}) becomes
\begin{equation} \label{rad}
(1-y)^4 f(y) R''(y) +(1-y)^4 f'(y)R'(y) + \left( \frac{\omega^2 r_H^2}{f(y)}- \ell(\ell+1) (1-y)^2 -m^2 r_H^2 \right) R(y)=0\, ,
\end{equation}
where the prime denotes derivative with respect to y. In the new coordinate the event horizon is located at $y=0$ and the spatial infinity at $y=1$. Now, we consider the boundary conditions. In the neighborhood of the horizon (y $\rightarrow$ 0) the function $R(y)$ behaves as
\begin{equation}
R(y)=C_1 e^{-\frac{i \omega r_H}{f'(0)} \ln{y}}+C_2 e^{\frac{i \omega r_H}{f'(0)} \ln{y}} \,.
\end{equation}
Here, the first term represents an ingoing wave and the second represents an outgoing wave near the black hole horizon. Imposing the requirement of only ingoing waves on the horizon, we fix $C_2=0$. On the other hand, at infinity the function $R(y)$ behaves as
\begin{equation}
R(y)= D_1 (1-y)^{\frac{3}{2} + \sqrt{\left(\frac{3}{2}  \right)^2 + m^2 R^2}}+ D_2 (1-y)^{\frac{3}{2} - \sqrt{\left(\frac{3}{2}  \right)^2 + m^2 R^2}} \,.
\end{equation}
So, imposing the scalar field vanishes at infinity requires $D_2=0$.
Taking into account the above behaviors of the scalar field at the horizon and at spatial infinity we define
\begin{equation}
R(y)= e^{-\frac{i \omega r_H}{f'(0)} \ln{y}} (1-y)^{\frac{3}{2} + \sqrt{\left(\frac{3}{2}  \right)^2 + m^2 R^2}} F(y) \,.
\end{equation}
Then, by inserting this last expression in Eq. (\ref{rad}) we obtain an equation for the function $F(y)$, which we solve numerically employing the pseudospectral Chebyshev method. The solution for the function $F(y)$ is assumed to be a finite linear combination of the Chebyshev polynomials, and it is inserted in the differential equation for $F(y)$. The interval $[0,1]$ is discretized at the Chebyshev collocation points. Then, the differential equation is evaluated at each collocation point. So, a system of algebraic equations is obtained, which corresponds to a generalized eigenvalue problem and it is solved numerically for $\omega$.

\begin{figure}[h!] 
\begin{center}  
\includegraphics[width=0.41\textwidth]{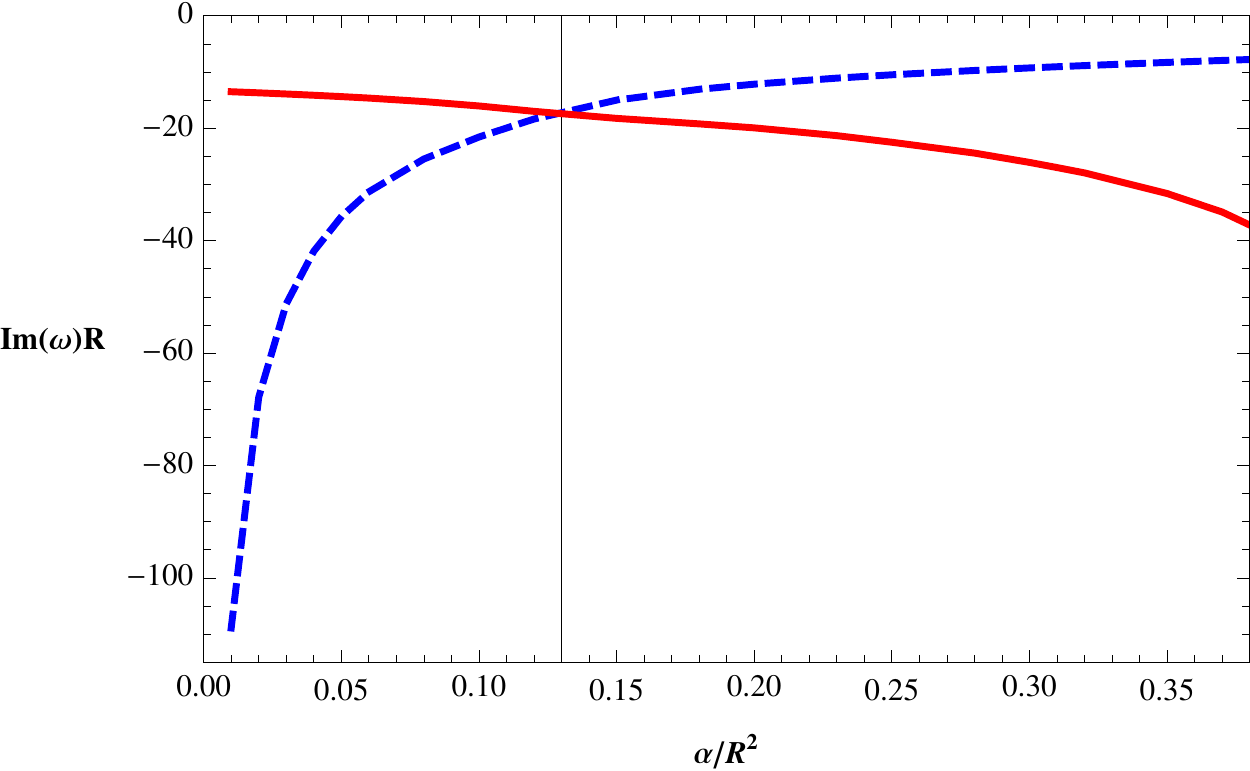}
\includegraphics[width=0.4\textwidth]{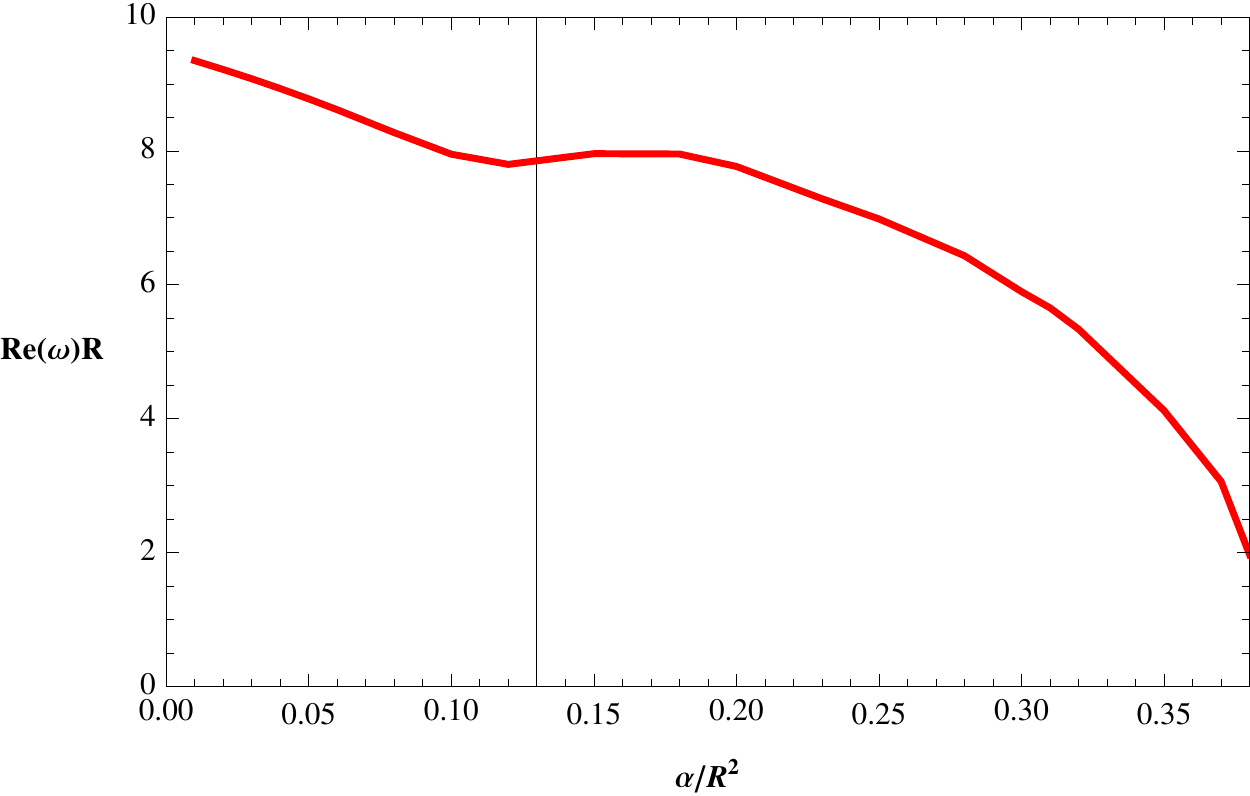}
\end{center}
\caption{Left panel for the behavior of the imaginary part of the QNFs for the nonperturbative in $\alpha$ modes (blue dashed line) and perturbative modes (red continuous line) of a massless scalar field. The vertical line corresponds to the critical value of $\alpha$ where the curves cross. 
Right panel for the behavior of the real part of QNFs for the perturbative modes of a massless scalar field, with $\ell =0$ as a function of $\alpha/R^2$, for
$r_H/R=5$.
} 
\label{F4}
\end{figure}

In Fig. \ref{F4}, left panel, we show the behavior of the imaginary part of the QNFs for a massless scalar field with $\ell=0$ as a function of $\alpha/R^2$, for 
a ratio $r_H/R=5$, and also
we show the real part of the QNFs, under the same considerations, right panel. We can observe the existence of two branches, one of them, corresponds to the branch perturbative in $\alpha$ (red continuous line) which consists of complex QNFs that in the limit $\alpha \rightarrow 0$, approximate to the QNFs of a massless scalar field in the background of the SAdS black hole, see \cite{Horowitz:1999jd}. On the other hand, the branch nonperturbative in $\alpha$ (blue dashed line) consists of purely imaginary QNFs, that diverge in the limit $\alpha \rightarrow 0$, therefore they do not exist for SAdS black hole. We show in Table \ref{T0} some numerical values of the QNFs. Also, we observe that the imaginary part of the QNFs is always negative for both branches; therefore, the propagation of massless scalar fields is stable in this background. It is worth to mention that there is a critical value $\alpha=\alpha_c$, where the curves intersect and both branches have the same imaginary part, for $\alpha$ lower than the critical value the nonperturbative branch decays faster than the perturbative branch, while that for $\alpha$ greater than the critical value, the behavior is opposite, i.e., the pertubative branch decays faster than the nonperturbative branch, thus the nonperturbative branch dominates in this case. The real part of the perturbative QNFs, see Fig. \ref{F4}, shows a smooth behavior and we observe that the frequency of the oscillation decreases when $\alpha/R^2$ increases, in addition, 
we observe that there is a small range where the frequency increases slightly and then decreases again. 

\begin{figure}[h!] 
\begin{center}  
\includegraphics[width=0.6\textwidth]{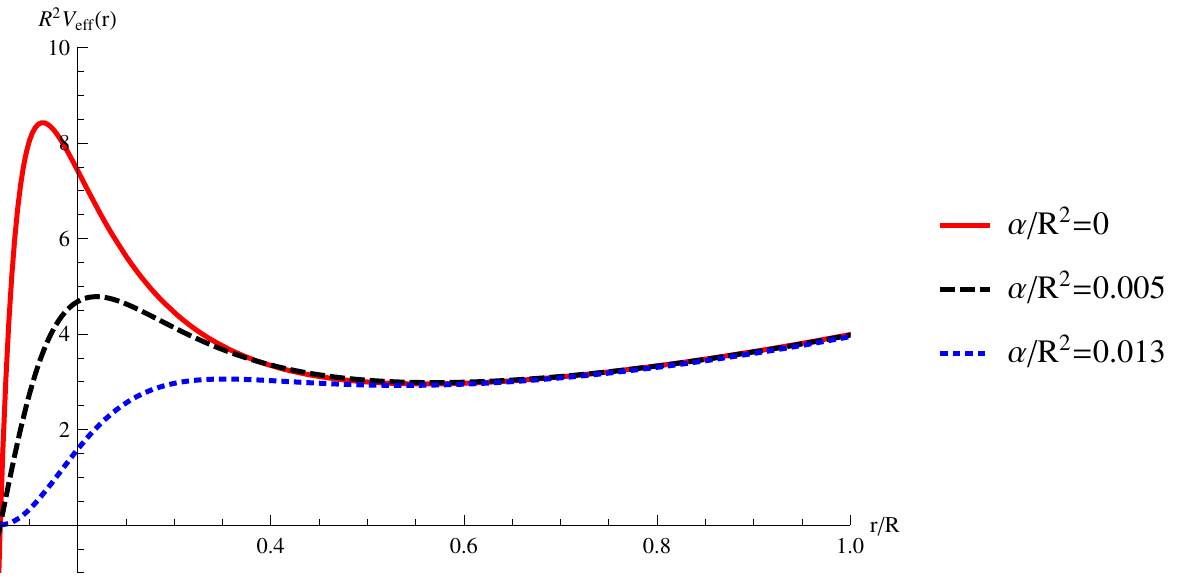}
\end{center}
\caption{The effective potential for small black holes for $r_H/R=0.12$, $mR=0$, $\ell=0$, and different values of $\alpha/R$.} 
\label{potenc}
\end{figure}

 We observe that for $\ell=m=0$
there exist two different potentials \footnote{We thank the referee  for pointing out this behaviour of the potential to us.} , one that looks like a potential barrier near the outside horizon-well-increasing, see Fig. \ref{potenc}, while the other is a monotonically increasing function like Fig. \ref{Potential}.
The former shows a feature of small SAdS black hole, whereas the latter indicates a large SAdS black hole. In \cite{Myung:2008pr} was showed that a potential-step type provides the purely imaginary QNFs, while the potential-barrier type gives the complex QNFs of a scalar field for the charged dilaton black hole. The presence of the bump near the horizon explains clearly why the QNFs for gravitational and electromagnetic perturbations of the small SAdS black hole are complex in \cite{Cardoso:2001bb}. For the 4D Einstein-Gauss-Bonnet AdS black hole, we observe a similar behavior of the potential, for small black holes and small values of $\alpha$ we note the presence of a potential barrier, which disappears when $r_H/R$ or $\alpha$ increases, see Fig. \ref{potenc}. Thus, it is possible to explain the two kinds of QNFs of a scalar field around the 4D Einstein-Gauss-Bonnet AdS black holes by identifying their potentials, i.e., while the potential-barrier type gives the complex QNFs, the monotonically increasing type gives the purely imaginary QNFs.
On the other hand, as mentioned, in Fig. \ref{F4} we observe that for $\alpha < \alpha_c$ the complex QNFs dominate, while that for $\alpha > \alpha_c$ the purely imaginary QNFs dominate. Interestingly, we found that this behavior is related to the change of concavity of the potential at the event horizon. We found that for $\alpha=0$ the second derivative of the effective potential evaluated at the horizon is always negative, and it is given by
 \begin{equation}
V_{eff}''(r_H)=- \frac{6(5+2 \ell (\ell+1))(r_H/R)^2+18(r_H/R)^4+2(4+3 \ell (\ell+1)+m^2 R^2 (r_H/R)^2))}{r_H^4}\,,
 \end{equation}
however, for $\alpha \neq 0$, the concavity of the potential at the event horizon can be positive, and we note that the potential has a point of inflection, see Fig. \ref{pot}, at the event horizon 
for $\alpha= \alpha_c$ where the curves in Fig. \ref{F4} intersect. For $\alpha < \alpha_c$ the potential has negative concavity at the event horizon, such as for the SAdS black hole, and the complex QNF dominates, while that for $\alpha > \alpha_c$ the concavity of the potential at the event horizon is positive and the purely imaginary QNF dominates. 
The change of the sign of $V_{eff}''(r_H)$ when $\alpha$ increases is attributed to the effect of the higher order curvature terms on the metric.


\begin{table}[ht]
\caption{Some lowest quasinormal frequencies $\omega R$ for the branches nonperturbative and perturbative in $\alpha$, in the background of the black hole with $r_H/R=5$, $\ell=0$ and $m=0$.}
\label{T0}\centering
\begin{tabular}{ | c | c | c |}
\hline
$\alpha/R^2$ & Nonperturbative QNFs  & Perturbative QNFs \\ \hline
$0.01$ &  $-108.89879 i$ & $9.34676- 13.50717i$  \\
$0.03$ &  $-51.24661 i$ &  $9.07679- 13.90768i$  \\
$0.06$  & $-31.31146 i$ & $8.61373 -14.63611 i$  \\
$0.15$ & $-14.99874 i$  &  $7.95765- 18.24388i$ \\ \hline
\end{tabular}%
\end{table}

\begin{figure}[h!] 
\begin{center}  
\includegraphics[width=0.6\textwidth]{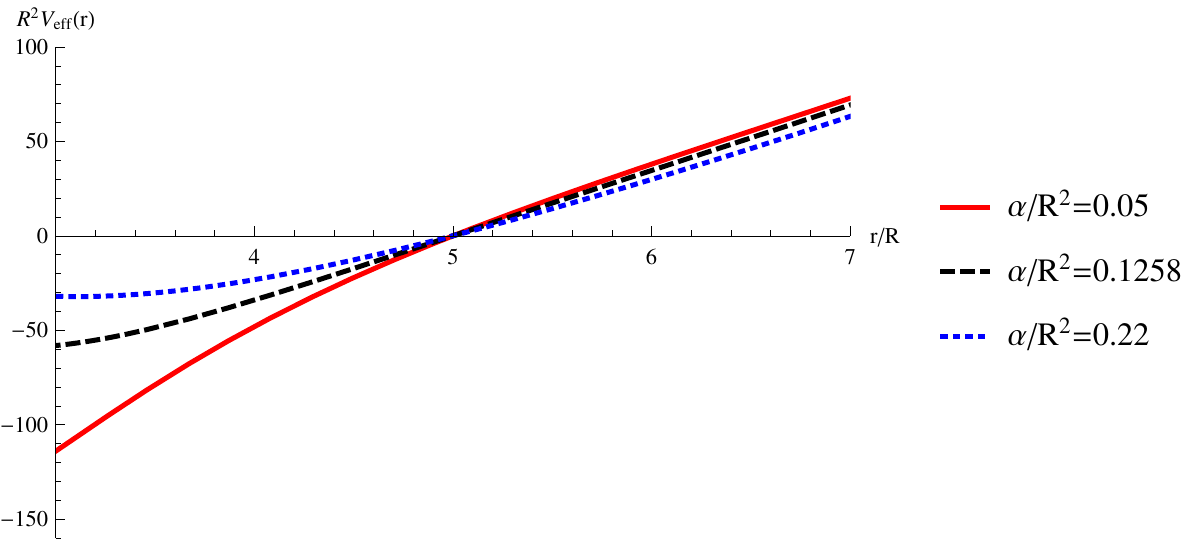}
\end{center}
\caption{The effective potential for $r_H/R=5$, $mR=0$, $\ell=0$ and $\alpha/R^2=0.05 < \alpha_c/R^2$, $\alpha/R^2=\alpha_c/R^2=0.1258$ and $\alpha/R^2=0.22>\alpha_c/R^2$.
} 
\label{pot}
\end{figure}

\begin{figure}[h!]
\begin{center}
\includegraphics[width=0.42\textwidth]{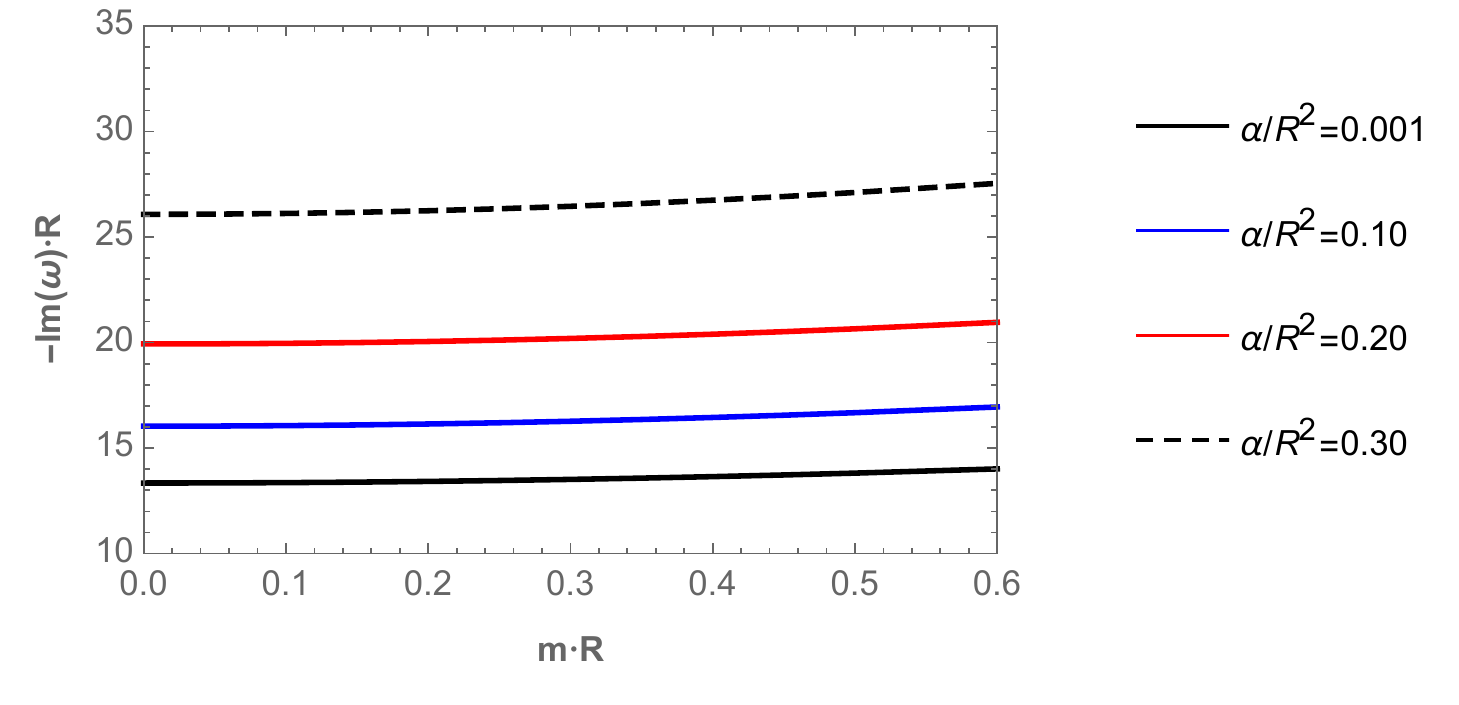}
\includegraphics[width=0.42\textwidth]{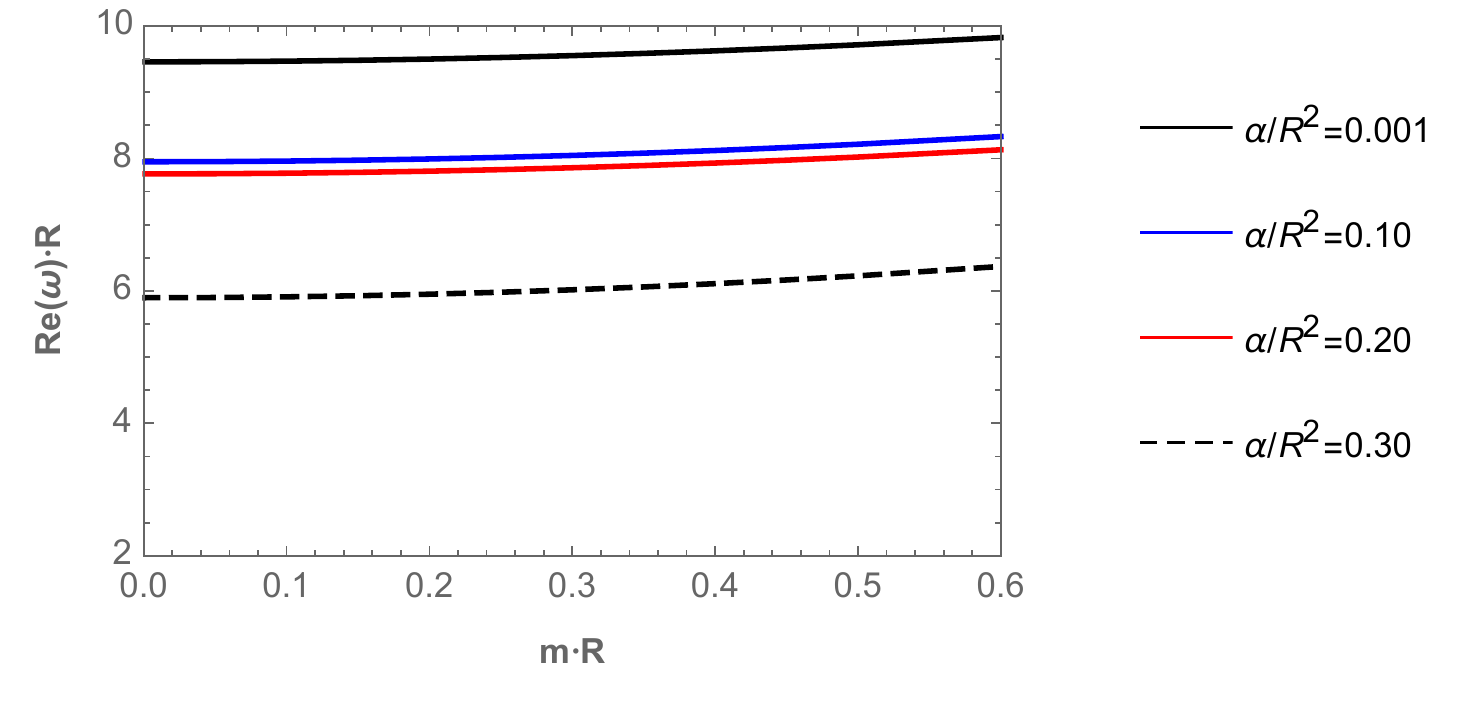}
\includegraphics[width=0.42\textwidth]{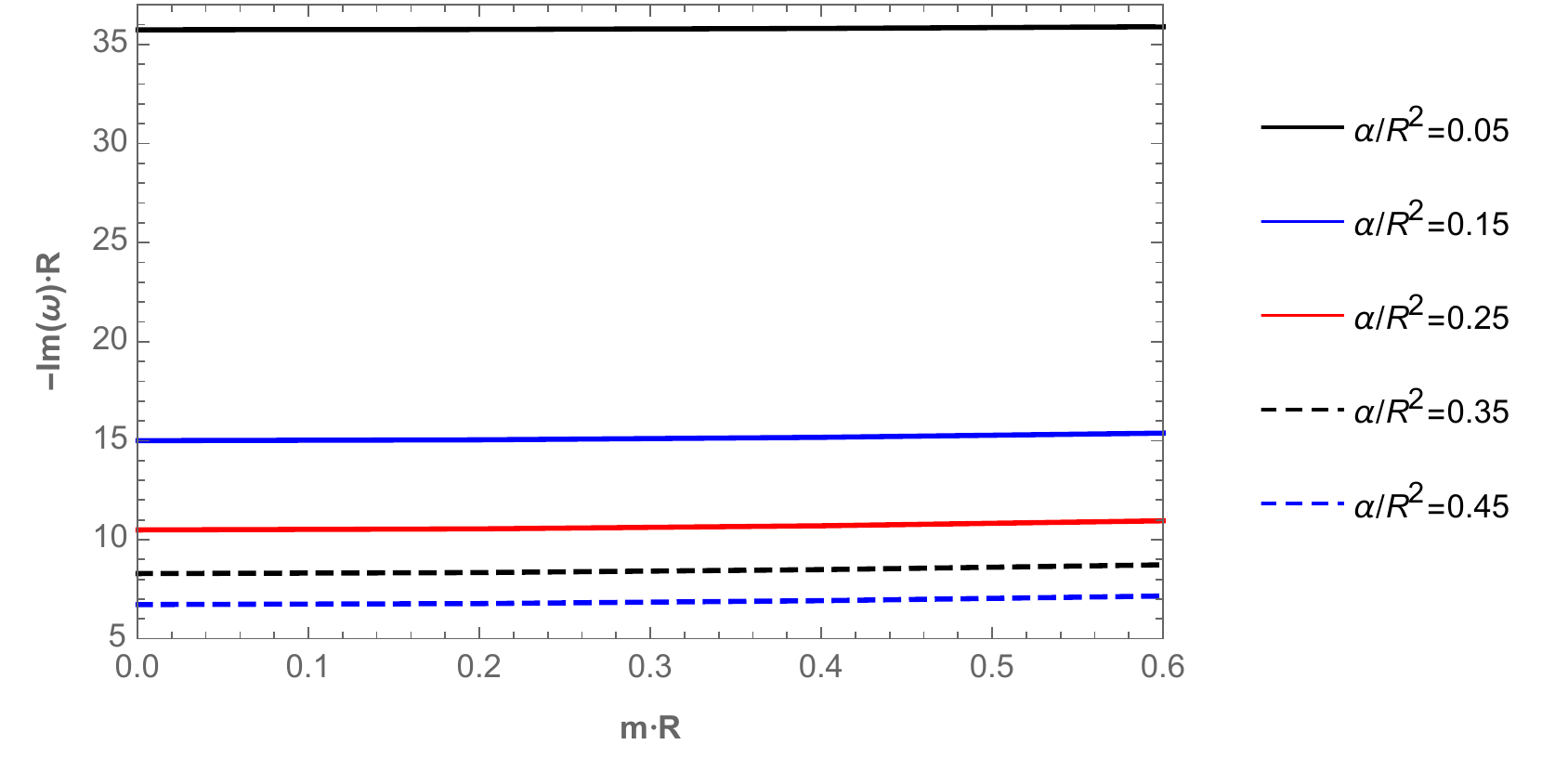}
\end{center}
\caption{The behavior of the QNFs as a function of $m R$, of a massive scalar field with $\ell =0$ for different values of $\alpha/R^2$ and $r_H/R=5$. Perturbative in $\alpha$ modes (top panel), left plot for the imaginary part and right plot for the real part; and nonperturbative modes (bottom panel).} \label{F5}
\end{figure}

Now, in order to analyze the behavior of the QNFs of massive scalar field, we plot in Fig. \ref{F5}, their behavior, for the lowest angular number $\ell=0$ as a function of $mR$, and for different values of $\alpha/R^2$. We can observe 
the complex branch (top panel) and the 
purely imaginary branch (bottom panel), which belong to the perturbative and nonperturbative branches, respectively. For the perturbative branch, we can observe that there is a faster decay rate of the perturbations when the mass of the scalar field increases, and the frequency of the oscillations increases too. Also, the decay rate and the frequency of the oscillation increases when $\alpha/R^2$ decreases, for a fixed value of $mR$.
On the other hand, for the nonperturbative branch the decay rate increases slightly when the scalar field mass increases. Also, the there is a faster decay when $\alpha/R^2$ decreases.

\begin{figure}[h]
\begin{center}
\includegraphics[width=0.42\textwidth]{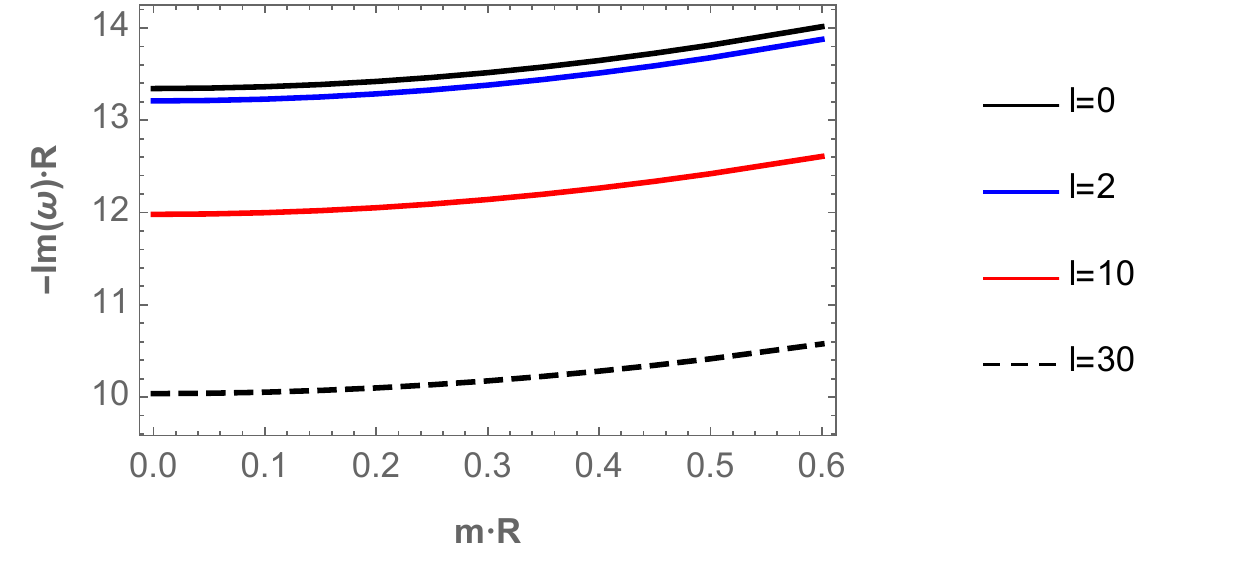}
\includegraphics[width=0.42\textwidth]{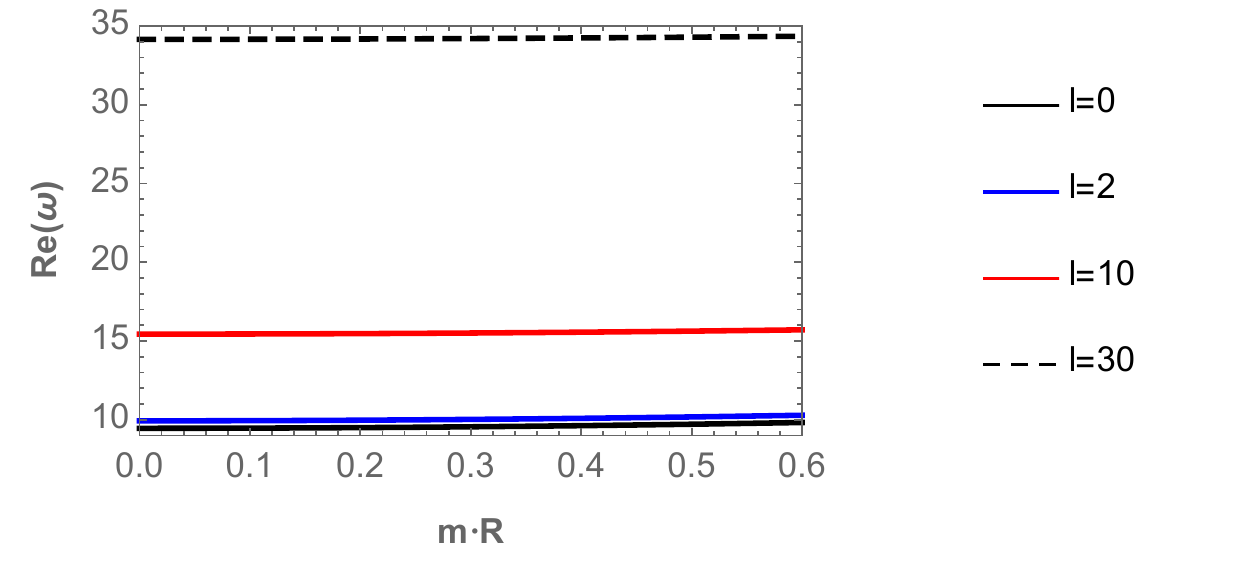}
\end{center}
\caption{The behavior of perturbative in $\alpha$ modes as a function of $m R$, with $r_H/R=5$, $\alpha/R^2=0.001$ and for different values of angular number $\ell =0,2,10,30$. Left plot for the imaginary part and right plot for the real part of the quasinormal spectrum.} \label{LB}
\end{figure}

\begin{figure}[h!]
\begin{center}
\includegraphics[width=0.42\textwidth]{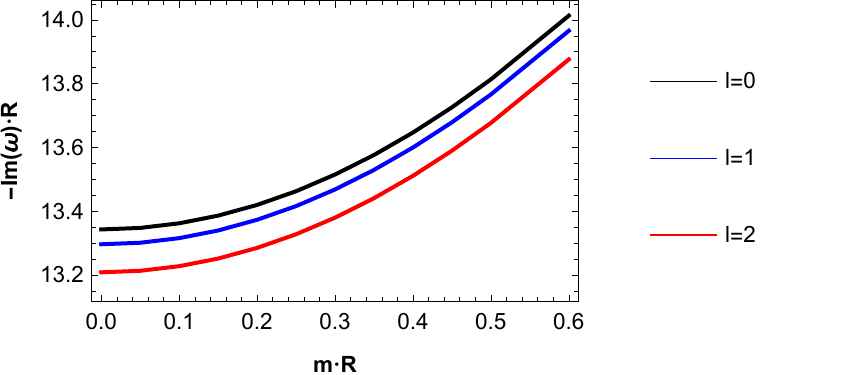}
\includegraphics[width=0.42\textwidth]{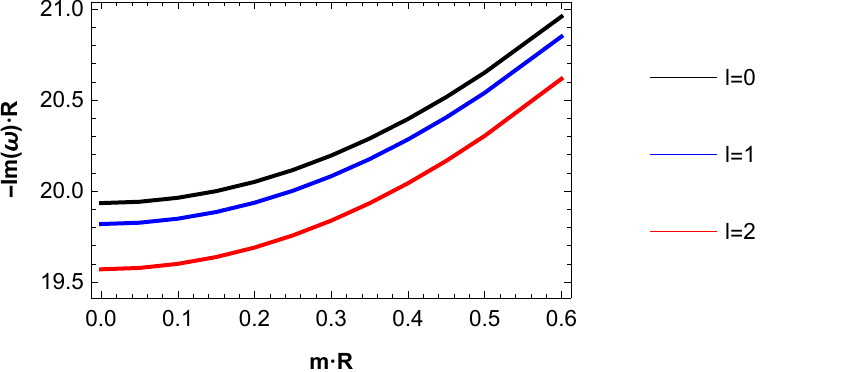}
\includegraphics[width=0.42\textwidth]{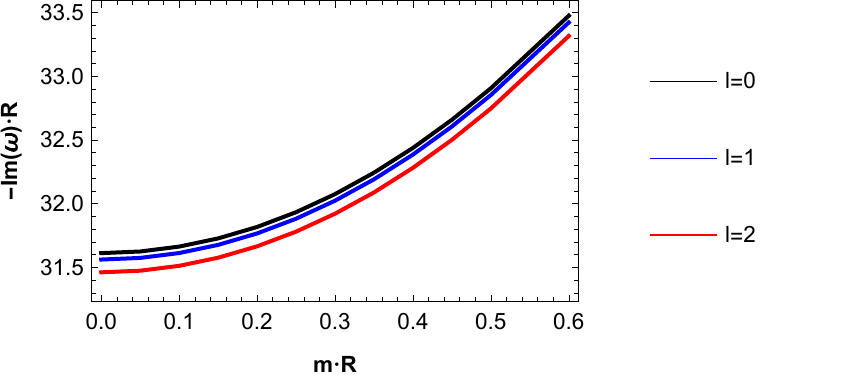}
\end{center}
\caption{The behavior of imaginary part of the perturbative in $\alpha$ modes as a function of $m R$, with $r_H/R=5$, for low values of angular number $\ell =0,1,2$. Left top panel for $\alpha/R^2=0.001$, right top panel for $\alpha/R^2=0.20$ and bottom panel for $\alpha/R^2=0.35$.} \label{F7}
\end{figure}

Now, in order to analyze the behavior of the QNFs of massive scalar field, we plot in Fig. \ref{LB}, their behavior, for low angular numbers $\ell=0,2$ and high angular numbers $\ell=10,30$ as a function of $mR$ with $\alpha/R^2=0.001$ fixed. For the perturbative branch, we can observe that there is a lower decay rate and the frequency of the oscillations increases when the angular number $\ell$ increases.
In Fig. \ref{F7}, we observe that the behavior is similar for low angular numbers $\ell=0,1,2$ and different values of $\alpha/R^2$.

\section{Conclusions}
\label{conclusion}

In this work, we considered 4D Einstein-Gauss-Bonnet black holes in AdS spacetime as backgrounds and we studied the propagation of probe scalar fields. We found numerically the quasinormal frequencies for different values of the Gauss-Bonnet coupling constant $\alpha /R^2$, the multipole number $\ell$ and the mass of the scalar field $m R$ by using the pseudospectral Chebyshev method. Mainly, we found two branches of QNFs, a branch perturbative in the coupling constant $\alpha$, and another branch nonperturbative in $\alpha$, that is, they do not exist in the limit $\alpha=0$. \\

The branch nonperturbative in $\alpha$ is characterized by purely imaginary QNFs with a faster decay when $\alpha/R^2$ decreases, while that for the branch perturbative in $\alpha$ the QNFs tend to the QNFs of the Schwarzschild AdS black hole when $\alpha \rightarrow 0$, and a lower decay is observed when $\alpha/R^2$ decreases. Also, we found that the imaginary part of the QNFs is always negative for both branches; therefore, the propagation of scalar fields is stable in the asymptotically AdS 4D Einstein-Gauss-Bonnet black hole. There are two different behaviors of potential, one that looks like a potential barrier near the outside horizon-well-increasing, while the other is a monotonically increasing function. The former shows a feature of small SAdS black hole, whereas the latter indicates a large SAdS black hole. For small black holes and small values of $\alpha$ we note the presence of a potential barrier, which disappears when $r_H/R$ or $\alpha$ increases. Therefore, it is posible to explain the two kinds of QNFs of a scalar field around the 4D Einstein-Gauss-Bonnet AdS black holes by identifying their potentials, i.e., while the potential-barrier type gives the complex QNFs, the monotonically increasing type gives the purely imaginary QNFs.

Interestingly, we found that there is a critical value of $\alpha =\alpha_c$, where both branches have the same imaginary part, and for values of $\alpha$ lower than the critical value the nonperturbative branch decays faster than the perturbative branch, while that for values of $\alpha/R^2$ greater than the critical value, the behavior is opposite, i.e., the pertubative branch decays faster than the nonperturbative branch, thus the nonperturbative branch dominates in this case. Additionally, we have found that for $\alpha=0$ the second derivative of the effective potential evaluated at the horizon is always negative, while that for $\alpha \neq 0$, the concavity of the potential at the event horizon can be positive, and we note that the potential has a point of inflection at the event horizon for $\alpha= \alpha_c$. For $\alpha < \alpha_c$ the potential has negative concavity at the event horizon, such as for the SAdS black hole, and the complex QNF dominates, while that for $\alpha > \alpha_c$ the concavity of the potential at the event horizon is positive and the purely imaginary QNF dominates. The change of sign of the second derivative of the effective potential evaluated at the horizon when $\alpha$ increases is attributed to the effect of the higher order curvature terms on the metric.
\\

We showed that the phenomena of nonperturbative modes arises for scalar field perturbations for the 4D Einstein-Gauss-Bonnet theory, by extending the presence of nonperturbative modes to other theory. On the other hand, the inverse of the imaginary part of the fundamental quasinormal frequency is related, through the AdS/CFT duality, to the thermalization time of the quantum states in the boundary field theory \cite{Horowitz:1999jd}. In addition, it was found in \cite{Grozdanov:2016fkt, Grozdanov:2016vgg} that black holes with AdS asymptotics in theories with higher curvature terms can help to describe the intermediate t'Hooft coupling in the dual field theory ; thus, we hope that the results obtained in this work can have applications along this line. \\

\acknowledgments
We would like to thank  the referee for his/her careful review of the manuscript and his/her valuable comments and suggestions which helped us to improve the manuscript considerably. Y.V. acknowledge support by the Direcci\'on de Investigaci\'on y Desarrollo de la Universidad de La Serena, Grant No. PR18142.


\begin{thebibliography}{11}
\bibliographystyle{prd}



\bibitem{Glavan:2019inb} 
  D.~Glavan and C.~Lin,
  Phys.\ Rev.\ Lett.\  {\bf 124}, no. 8, 081301 (2020)
  [arXiv:1905.03601 [gr-qc]].
  
  \bibitem{Fernandes:2020rpa} 
  P.~G.~S.~Fernandes,
  arXiv:2003.05491 [gr-qc].
  
  \bibitem{Ghosh:2020syx}
  S.~G.~Ghosh and R.~Kumar,
  arXiv:2003.12291 [gr-qc].
  
  \bibitem{Singh:2020nwo} 
  D.~V.~Singh, S.~G.~Ghosh and S.~D.~Maharaj,
  arXiv:2003.14136 [gr-qc].

\bibitem{Wei:2020ght} 
  S.~W.~Wei and Y.~X.~Liu,
  arXiv:2003.07769 [gr-qc].


  \bibitem{Kumar:2020uyz} 
  A.~Kumar and R.~Kumar,
  arXiv:2003.13104 [gr-qc].
  

  

\bibitem{Konoplya:2020ibi} 
  R.~A.~Konoplya and A.~Zhidenko,
  arXiv:2003.12171 [gr-qc].
  
    
\bibitem{Konoplya:2020qqh} 
  R.~A.~Konoplya and A.~Zhidenko,
  Phys.\ Rev.\ D {\bf 101}, no. 8, 084038 (2020)
  [arXiv:2003.07788 [gr-qc]].

\bibitem{Konoplya:2020der} 
  R.~A.~Konoplya and A.~Zhidenko,
  arXiv:2005.02225 [gr-qc].

 \bibitem{Hegde:2020xlv} 
  K.~Hegde, A.~Naveena Kumara, C.~L.~A.~Rizwan, A.~K.~M. and M.~S.~Ali,
  arXiv:2003.08778 [gr-qc].
 
 \bibitem{HosseiniMansoori:2020yfj} 
  S.~A.~Hosseini Mansoori,
  arXiv:2003.13382 [gr-qc].
 
  \bibitem{Wei:2020poh} 
  S.~W.~Wei and Y.~X.~Liu,
  arXiv:2003.14275 [gr-qc].
  
  \bibitem{Singh:2020xju}
  D.~V.~Singh and S.~Siwach,
  arXiv:2003.11754 [gr-qc].
  
  \bibitem{EslamPanah:2020hoj} 
  B.~Eslam Panah and K.~Jafarzade,
  arXiv:2004.04058 [hep-th].
  
  

\bibitem{Zhang:2020qam} 
  C.~Y.~Zhang, P.~C.~Li and M.~Guo,
  arXiv:2003.13068 [hep-th].
  
\bibitem{Konoplya:2020cbv} 
  R.~A.~Konoplya and A.~F.~Zinhailo,
  arXiv:2004.02248 [gr-qc].
  
  
  \bibitem{Konoplya:2020bxa} 
  R.~A.~Konoplya and A.~F.~Zinhailo,
  arXiv:2003.01188 [gr-qc].
  
 
  
  \bibitem{Konoplya:2020juj} 
  R.~A.~Konoplya and A.~Zhidenko,
  arXiv:2003.12492 [gr-qc].
  
  \bibitem{Mishra:2020gce} 
  A.~K.~Mishra,
  arXiv:2004.01243 [gr-qc].
  
  \bibitem{Churilova:2020aca} 
  M.~S.~Churilova,
  arXiv:2004.00513 [gr-qc].
  
  \bibitem{Zhang:2020sjh} 
  C.~Y.~Zhang, S.~J.~Zhang, P.~C.~Li and M.~Guo,
  arXiv:2004.03141 [gr-qc].
  
  \bibitem{Guo:2020zmf} 
  M.~Guo and P.~C.~Li,
  arXiv:2003.02523 [gr-qc].
 
 \bibitem{Heydari-Fard:2020sib}
M.~Heydari-Fard, M.~Heydari-Fard and H.~Sepangi,
[arXiv:2004.02140 [gr-qc]]


 \bibitem{Roy:2020dyy}
R.~Roy and S.~Chakrabarti,
[arXiv:2003.14107 [gr-qc]].
  
  \bibitem{Liu:2020vkh} 
  C.~Liu, T.~Zhu and Q.~Wu,
  arXiv:2004.01662 [gr-qc].





\bibitem{Gurses:2020ofy} 
  M.~Gurses, T.~C.~Sisman and B.~Tekin,
  arXiv:2004.03390 [gr-qc].
  
  
  \bibitem{Mahapatra:2020rds} 
  S.~Mahapatra,
  arXiv:2004.09214 [gr-qc].

\bibitem{Shu:2020cjw} 
  F.~W.~Shu,
  arXiv:2004.09339 [gr-qc].
  
  \bibitem{Tian:2020nzb} 
  S.~X.~Tian and Z.~H.~Zhu,
  arXiv:2004.09954 [gr-qc].
  
  \bibitem{Lu:2020iav} 
  H.~Lu and Y.~Pang,
  arXiv:2003.11552 [gr-qc].
  
  
  \bibitem{Fernandes:2020nbq} 
  P.~G.~S.~Fernandes, P.~Carrilho, T.~Clifton and D.~J.~Mulryne,
  arXiv:2004.08362 [gr-qc].
  
  \bibitem{Hennigar:2020lsl} 
  R.~A.~Hennigar, D.~Kubiznak, R.~B.~Mann and C.~Pollack,
  arXiv:2004.09472 [gr-qc].
  
  
  \bibitem{Mann:1992ar} 
  R.~B.~Mann and S.~F.~Ross,
  Class.\ Quant.\ Grav.\  {\bf 10}, 1405 (1993)
  [gr-qc/9208004].
  
  
  \bibitem{Kobayashi:2020wqy} 
  T.~Kobayashi,
  arXiv:2003.12771 [gr-qc].



\bibitem{Abbott:2016blz}
  B.~P.~Abbott {\it et al.} [LIGO Scientific and Virgo Collaborations],
  Phys.\ Rev.\ Lett.\  {\bf 116}, no. 6, 061102 (2016)

  
\bibitem{Regge:1957td} 
  T.~Regge and J.~A.~Wheeler,
  Phys.\ Rev.\  {\bf 108}, 1063 (1957).
  
  
\bibitem{Zerilli:1971wd} 
  F.~J.~Zerilli,
  Phys.\ Rev.\ D {\bf 2}, 2141 (1970).
  
  
\bibitem{Kokkotas:1999bd} 
  K.~D.~Kokkotas and B.~G.~Schmidt,
  Living Rev.\ Rel.\  {\bf 2}, 2 (1999)
  [gr-qc/9909058].
  
  
\bibitem{Nollert:1999ji} 
  H.~P.~Nollert,
  Class.\ Quant.\ Grav.\  {\bf 16}, R159 (1999).
  
  
\bibitem{Konoplya:2011qq}
  R.~A.~Konoplya and A.~Zhidenko,
  Rev.\ Mod.\ Phys.\  {\bf 83}, 793 (2011)
  [arXiv:1102.4014 [gr-qc]];
  
  
  E.~Berti, V.~Cardoso and A.~O.~Starinets,
  Class.\ Quant.\ Grav.\  {\bf 26}, 163001 (2009)
  [arXiv:0905.2975 [gr-qc]];
  K.~D.~Kokkotas and B.~G.~Schmidt,
  Living Rev.\ Rel.\  {\bf 2}, 2 (1999)
  [gr-qc/9909058].

  
  
\bibitem{TheLIGOScientific:2016src}
  B.~P.~Abbott {\it et al.} [LIGO Scientific and Virgo Collaborations],
  Phys.\ Rev.\ Lett.\  {\bf 116}, no. 22, 221101 (2016)

\bibitem{Konoplya:2016pmh}
  R.~Konoplya and A.~Zhidenko,
  Phys.\ Lett.\ B {\bf 756}, 350 (2016)





 \bibitem{Gonzalez:2017gwa} 
  P.~A.~Gonzalez, R.~A.~Konoplya and Y.~Vasquez,
  Phys.\ Rev.\ D {\bf 95}, no. 12, 124012 (2017)
  [arXiv:1703.06215 [gr-qc]].


 



 \bibitem{Grozdanov:2016vgg}
  S.~Grozdanov, N.~Kaplis and A.~O.~Starinets,
  JHEP {\bf 1607}, 151 (2016)
  [arXiv:1605.02173 [hep-th]].
 

\bibitem{Konoplya:2017ymp} 
  R.~A.~Konoplya and A.~Zhidenko,
  Phys.\ Rev.\ D {\bf 95}, no. 10, 104005 (2017)
  [arXiv:1701.01652 [hep-th]].

\bibitem{Gonzalez:2018xrq} 
  P.~A.~Gonzalez, Y.~Vasquez and R.~N.~Villalobos,
  Phys.\ Rev.\ D {\bf 98}, no. 6, 064030 (2018)
  [arXiv:1807.11827 [gr-qc]].

  
    
 \bibitem{Grozdanov:2016fkt}
  S.~Grozdanov and A.~O.~Starinets,
  arXiv:1611.07053 [hep-th].
  
\bibitem{Takahashi:2010ye} 
  T.~Takahashi and J.~Soda,
  Prog.\ Theor.\ Phys.\  {\bf 124}, 911 (2010)
  [arXiv:1008.1385 [gr-qc]].


\bibitem{Dotti:2005sq}
  G.~Dotti, R.~J.~Gleiser,
  Phys.\ Rev.\ D {\bf 72}, 044018 (2005)
  [gr-qc/0503117].


\bibitem{Gleiser:2005ra}
  R.~J.~Gleiser, G.~Dotti,
  Phys.\ Rev.\ D {\bf 72}, 124002 (2005)
  [gr-qc/0510069].



  

\bibitem{Boyd}
J. P. Boyd, Chebyshev and Fourier Spectral Methods. Dover Books on Mathematics. Dover Publications, Mineola, NY, second ed., 2001.
  
  
\bibitem{Finazzo:2016psx} 
  S.~I.~Finazzo, R.~Rougemont, M.~Zaniboni, R.~Critelli and J.~Noronha,
  JHEP {\bf 1701}, 137 (2017)
  [arXiv:1610.01519 [hep-th]].
  
\bibitem{Gonzalez:2017shu} 
  P.~A.~González, E.~Papantonopoulos, J.~Saavedra and Y.~Vásquez,
  Phys.\ Rev.\ D {\bf 95}, no. 6, 064046 (2017)
  [arXiv:1702.00439 [gr-qc]].



\bibitem{Cai:2009ua} 
  R.~G.~Cai, L.~M.~Cao and N.~Ohta,
  JHEP {\bf 1004}, 082 (2010)
  [arXiv:0911.4379 [hep-th]].
  
  

\bibitem{Cognola:2013fva} 
  G.~Cognola, R.~Myrzakulov, L.~Sebastiani and S.~Zerbini,
  Phys.\ Rev.\ D {\bf 88}, no. 2, 024006 (2013)
  [arXiv:1304.1878 [gr-qc]].
  
  
\bibitem{Horowitz:1999jd} 
  G.~T.~Horowitz and V.~E.~Hubeny,
  Phys.\ Rev.\ D {\bf 62}, 024027 (2000)
  [hep-th/9909056].


\bibitem{Myung:2008pr} 
  Y.~S.~Myung, Y.~W.~Kim and Y.~J.~Park,
  Eur.\ Phys.\ J.\ C {\bf 58}, 617 (2008)
  [arXiv:0809.1933 [gr-qc]].


\bibitem{Cardoso:2001bb} 
  V.~Cardoso and J.~P.~S.~Lemos,
  Phys.\ Rev.\ D {\bf 64}, 084017 (2001)
  [gr-qc/0105103].

\bibitem{Chan:1996yk}
J.~Chan and R.~B.~Mann,
Phys. Rev. D \textbf{55} (1997), 7546-7562
[arXiv:gr-qc/9612026 [gr-qc]].


  
\bibitem{Tomozawa:2011gp} 
  Y.~Tomozawa,
  arXiv:1107.1424 [gr-qc].
  


\end{thebibliography}
\end{document}